\documentclass[a4paper,accepted=2023-06-29]{quantumarticle}
\pdfoutput=1
\usepackage[numbers,sort&compress]{natbib}

\usepackage{amsmath,bbold,braket,graphicx,xcolor}
\usepackage{siunitx}[=v2]
\usepackage[version=4]{mhchem}

\usepackage[utf8]{luainputenc}

\graphicspath{{images/}}

\newcommand{\op}[1]{\ensuremath{\hat{#1}}}
\newcommand{\mat}[1]{\ensuremath{\mathbf{#1}}}

\newcommand{\dd}{\ensuremath{\mathrm{d}}}
\newcommand{\ee}{\ensuremath{\mathrm{e}}}
\newcommand{\ii}{\ensuremath{\mathrm{i}}}

\newcommand{\rootnot}{\ensuremath{\sqrt{\mathrm{NOT}}}}

\usepackage[hidelinks]{hyperref}

\begin{document}

\title{A quantum logic gate for free electrons}

\author[ustem]{Stefan Löffler}
\ead{stefan.loeffler@tuwien.ac.at}

\author[ustem,ifp]{Thomas Schachinger}

\author[ceos]{Peter Hartel}

\author[erc,rwth]{Peng-Han Lu}

\author[erc]{Rafal E. Dunin-Borkowski}

\author[kit]{Martin Obermair}

\author[kit]{Manuel Dries}

\author[kit]{Dagmar Gerthsen}

\author[ustem,ifp]{Peter Schattschneider}

\affiliation[ustem]{University Service Centre for Transmission Electron Microscopy, TU Wien, Wiedner Hauptstraße 8-10/E057-02, 1040 Wien, Austria}

\affiliation[ifp]{Institute of Solid State Physics, TU Wien, Wiedner Hauptstraße 8-10/E138-03, 1040 Wien, Austria}

\affiliation[ceos]{CEOS Corrected Electron Optical Systems GmbH, Englerstraße 28, 69126 Heidelberg, Germany}

\affiliation[erc]{Ernst Ruska-Centre for Microscopy and Spectroscopy with Electrons (ER-C) and Peter Grünberg Institute, Forschungszentrum Jülich, 52425 Jülich, Germany}

\affiliation[rwth]{RWTH Aachen University, Ahornstraße 55, 52074 Aachen, Germany}

\affiliation[kit]{Laboratorium für Elektronenmikroskopie (LEM), Karlsruher Institut für Technologie (KIT), Engesserstraße 7, 76131 Karlsruhe, Germany}

\begin{abstract}
The topological charge $m$ of vortex electrons spans an infinite-dimensional Hilbert space. Selecting a two-dimensional subspace spanned by $m=\pm 1$, a beam electron in a transmission electron microscope (TEM) can be considered as a quantum bit (qubit) freely propagating in the column. A combination of electron optical quadrupole lenses can serve as a universal device to manipulate such qubits at the experimenter's discretion. We set up a TEM probe forming lens system as a quantum gate and demonstrate its action numerically and experimentally. High-end TEMs with aberration correctors are a promising platform for such experiments, opening the way to study quantum logic gates in the electron microscope.
\end{abstract}

\maketitle

\section{Introduction}
Manipulating the electron's phase is a current topic in electron microscopy.
On the one hand, wave front engineering promises better spatial resolution \cite{PRA_v11_i_p44072}, novel beam splitters~\cite{Hommelhof2015}, improved sensitivity for particular applications such as spin polarized electronic transitions~\cite{Schachinger2017}, or manipulating nanoparticles via electron vortex beams~\cite{VerbeeckAdvMat2013}. In many respects, the physics of electrons with topological charge is similar to that of photons in singular optics (for an overview see~\cite{Franke2008}). In particular, quantum logic gates based on photons with orbital angular momentum (OAM) have been successfully demonstrated (e.g. in~\cite{Zeilinger2017}). Other aspects are unique to electrons, such as easy manipulation in magnetic fields, the extraordinary sub-nm resolution, or novel solid-state applications such as diffraction in chiral crystals~\cite{Juchtmans2015}. On the other hand, the coherent control of the interaction of fast electrons with electromagnetic radiation, either via near fields in PINEM~\cite{Carbone2018,Feist2015200}, resonant cavities~\cite{Baum2016} or laser accelerators~\cite{Hommelhoff2019} leads to oscillations in the probability distribution of the electron's momentum and energy, allowing the compression of fast electron pulses below the femtosecond time scale.

After the poineering work of Bliokh~\cite{Bliokh2007,Bliokh2011} on electron vortices it took some time for their experimental realization~\cite{Verbeeck2010,N_v464_i_p737}. It was soon shown that they can be manipulated in a magnetic field, giving rise to peculiar rotations~\cite{Bliokh2012,NC_v5_i_p4586,PRL_v110_i_p93601,U_v158_i_p17}. For a review see~\cite{Bliokh2017}. The possibility to shape the phase of the electron wave with special holographic masks or via interaction with electromagnetic fields allows  not only to prepare single electron wave packets propagating  in free space as qubits but also to implement quantum gates for such electrons. This opens the way to design a new platform for quantum operations. 

There are many such platforms available, each of them based on a different physical realization of the qubits. Perhaps the best known examples are entangled photon qubits which have been shown to be scalable to some 10$^4$ entangled modes~\cite{qubits-Larsen}; also trapped ion~\cite{qubits-Brown} or superconducting~\cite{qubits-Kjaergaard} qubits were demonstrated. Each of them offers particular benefits and disadavantages. For the time being, it is not clear which of the many experimentally demonstrated implementations will win the race in quantum computing. In recent years, solid state qubits demonstrated promising features in terms of fidelity, scalability or lifetime such as nitrogen-vacancy color centers in diamond (a 10-qubit register that can store quantum information for one minute has recently been demonstrated~\cite{qubits-Bradley}) or  spin qubits based on the well established silicon technology. For overviews, see, e.g., \cite{RoPiP_v74_i10_104401,qubits-Chatterjee}.

Recently, an  approach towards free electron qubits in the electron microscope, based on energy gain or loss processes using laser-driven near field interactions was proposed~\cite{Reinhardt2021,PRL_v126_i23_233403,PRR_v3_i4_43033}  but to our knowledge, quantum gates for their manipulation have not yet been realized.

Here, we present proof-of-principle experiments with a quantum $\rootnot$ gate for freely floating electrons. To this aim, we use a device designed for electron microscopes, called a mode converter (MC)~\cite{U_v234_i_113456} that transforms a plane electron wave into one with topological charge~\cite{Schattschneider2012,U_v229_i_113340}. It should be noted at this point that free floating electron qubits in a TEM are a very novel and emerging field of research. As such, they are not on the level of sophistication of, e.g., photonic qubits that have been researched for decades. In particular, this work deals with the production, manipulation and readout of single qubits. While further concepts such as multi-qubit manipulation and entanglement are briefly discussed, they are essentially beyond the scope of this paper and will require further investigation. Despite the fact that free floating electrons are very new to the scene of quantum information and computing, we strongly believe that they can provide novel insights, particularly in the realm of fundamental research.

\section{Theory}
\subsection{Basis states}
In a two-state system, any two orthogonal states can be chosen as a basis for constructing qubits. Preliminary experimental results show that vortex electrons --- eigenmodes of the angular momentum operator that are topologically protected and carry quantized OAM of integer multiples of $\hbar$ --- are very stable during manipulation in the column of a microscope~\cite{U_v229_i_113340}. 
Therefore, a Hilbert space spanned by two vortex states with topological charge $m= \pm 1$ (and linear combinations thereof) is a good candidate for electron qubits.

\begin{figure}
	\centering
	\includegraphics[width=\columnwidth]{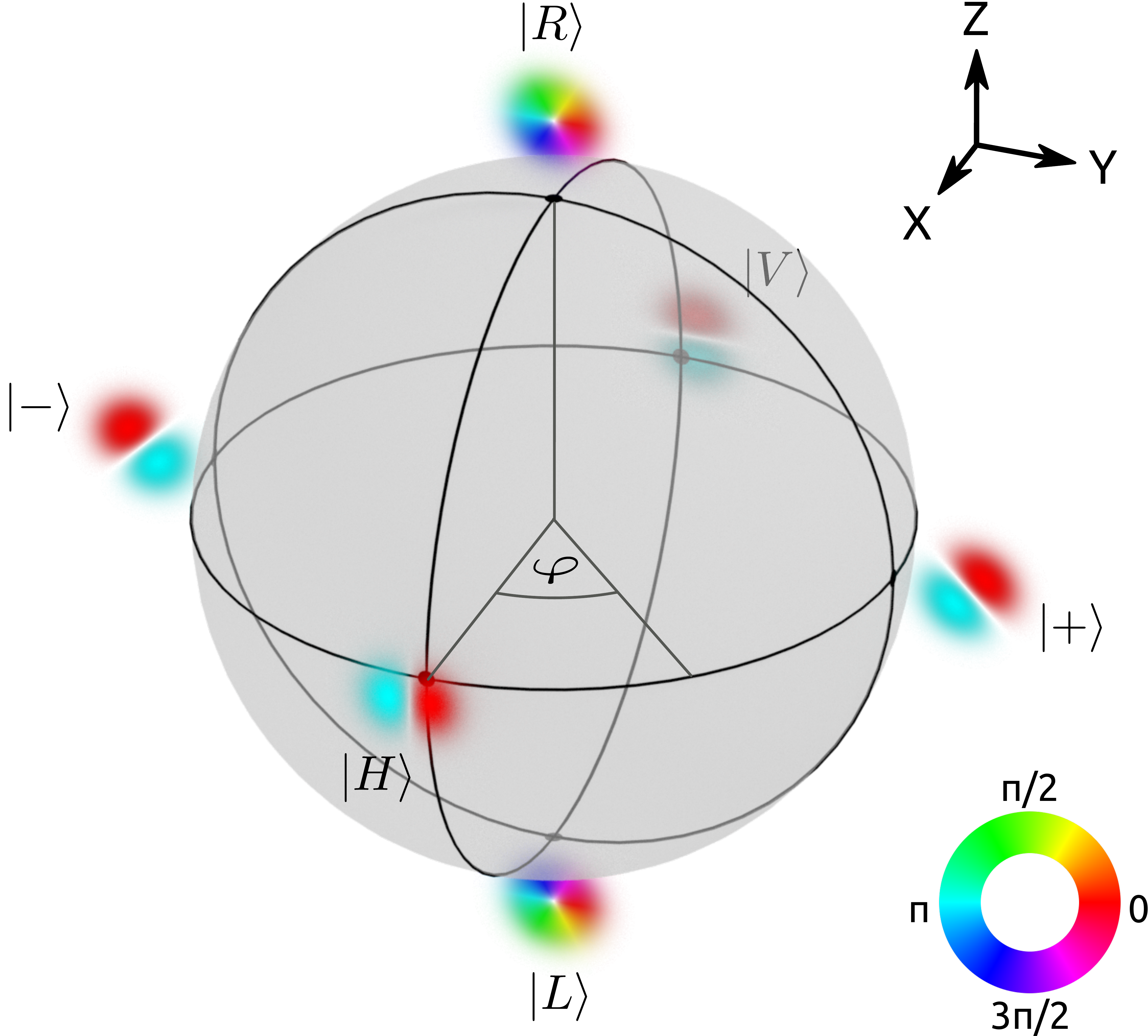}
	\caption{$\ket{R}$ and $\ket{L}$ states are represented by the north and south poles of the Bloch sphere. Also shown are the $\ket{\pm}$ and the $\ket{H},  \ket{V}$ states.
	}
	\label{fig:bloch-sphere}
\end{figure}

We use the two Laguerre-Gauss (LG) modes $LG_{1,0}$ and $LG_{-1,0}$ as basis states \cite{Bliokh2012}. In cylindrical coordinates $(r,\phi, z)$ they have the real-space representations
\begin{equation}
\begin{aligned}
\braket{\vec r \, | R} = LG_{1,0} &= r \ee^{\ii \phi} \cdot f(r,z), \\
\braket{\vec r \, | L} = LG_{-1,0} &= r \ee^{-\ii \phi} \cdot f(r,z)
\label{LGR}
\end{aligned}
\end{equation}
with
\[
f(r,z)=\frac{A}{w(z)}\ee^{-\frac{r^2}{w(z)^2}} \cdot \ee^{\ii \frac{k r^2}{2 R(z)}} \cdot \ee^{\ii (kz - 2\zeta(z))},
\]
where $A$ is a real valued normalization factor. The waist $w(z)=w_0 \sqrt{1+(z/z_R)^2}$ is the propagation dependent beam size, $z_R$ is the Rayleigh length, $k$ is the wave number, $R(z)=z(1+(z_R/z)^2)$ is the radius of curvature of the wave front, and $\zeta(z)=\arctan(z_R/z)$ is the Gouy phase~\footnote{It has been pointed out that for non-relativistic electrons the Gouy phase depends on the time at which the propagating wave packet is observed~\cite{Karlovets2018}, as $t=\braket{z}/v$ where $v$ is the electron's speed.}.
These are diffracting modes, i.e. the radial scale depends on the position $z$ of the wave packet along the propagation axis. Note that $z$ is considered here as a parameter used for propagation simulation. If not otherwise stated, the qubits are defined in the virtual or real focal planes $z=0$. The two-dimensional Hilbert space spanned by $\ket{R}$ and $\ket{L}$ is conveniently presented as a Bloch sphere (Fig.~\ref{fig:bloch-sphere}). Similarly to light optics, we define states 
\begin{equation}
\ket{H} = \frac{1}{\sqrt{2}} (\ket{R} + \ket{L}) \qquad
\ket{V} = \frac{1}{\sqrt{2}} (\ket{R} - \ket{L})
\label{HV}
\end{equation}
and
\begin{equation}
\ket{+} = \frac{1}{\sqrt{2}} (\ket{R} + \ii\ket{L}) \qquad
\ket{-} = \frac{1}{\sqrt{2}} (\ket{R} - \ii\ket{L})
\label{PM}
\end{equation}

\begin{figure}[htbp]
	\centering
		\includegraphics[width=\columnwidth]{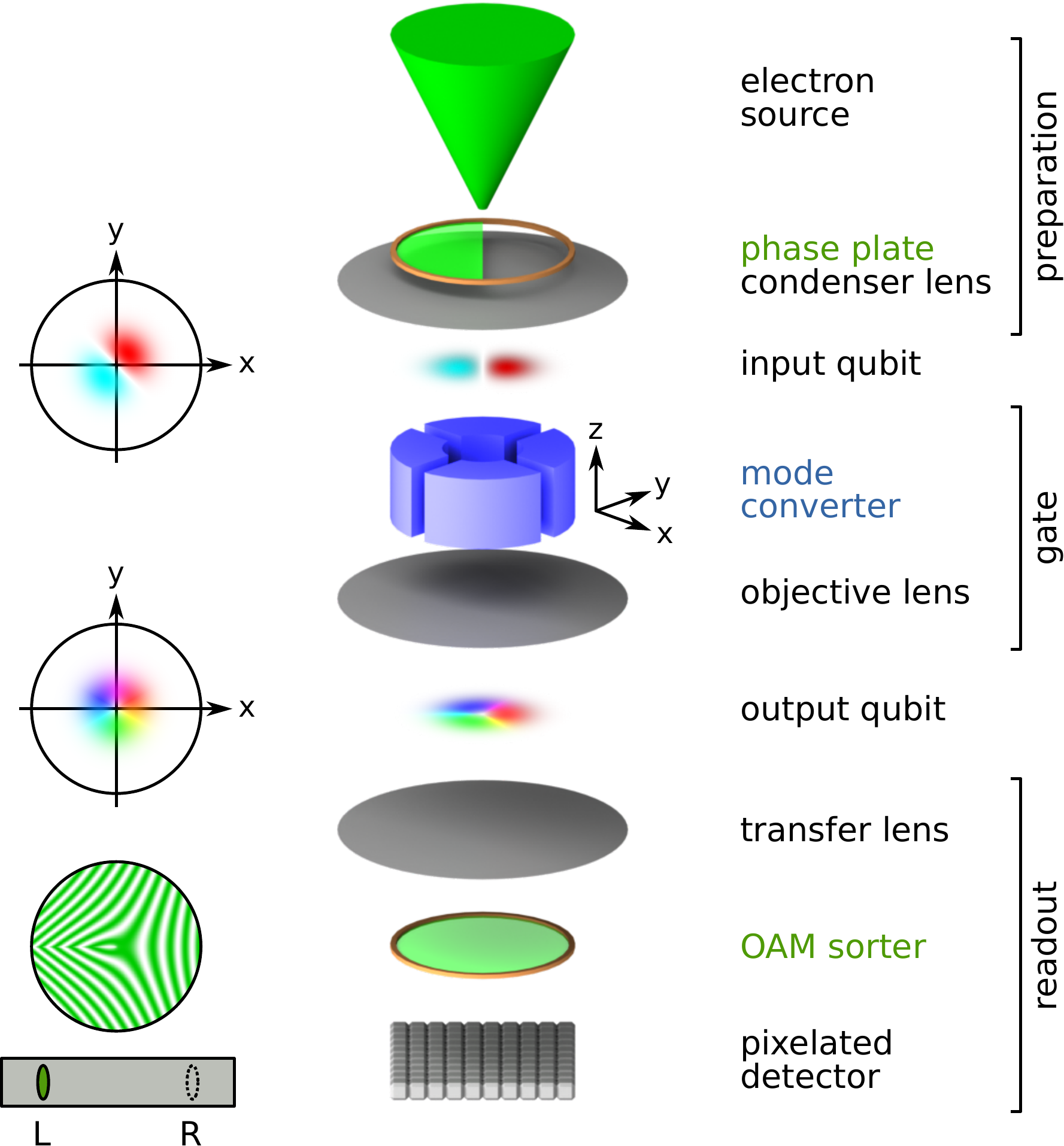}
	\caption{Column of an electron microscope with standard devices (black), phase shaping devices (green) and the qubit manipulator (blue). Electron qubits (color coded as in Fig.~\ref{fig:bloch-sphere}) travel down the $z$ axis.
}
	\label{fig:column}
\end{figure}

Performing qubit operations using a quantum logic gate requires three steps: preparation, manipulation using the gate, and readout, as sketched in Fig.~\ref{fig:column}.

\subsection{Qubit preparation}

For preparing the input qubit, the electron beam is sent through a phase plate. For the proof-of-principle experiment, we prepare input qubits as states on the equator of the Bloch sphere
\begin{equation}
\ket{I_\varphi} = \frac{1}{\sqrt{2}} ( \ket{R} + \ee^{\ii \varphi} \ket{L} ) 
\label{eq:Iphi}
\end{equation}
with phase shift $\varphi \in [0, \, 2 \pi)$.
Using Eq.~\ref{LGR},
\begin{equation}
\braket{\vec r \,|I_\varphi}=\ee^{\ii \varphi/2} \sqrt{2} \, f(r,z)\, r \cos(\phi-\varphi/2).
\label{eq:IR}
\end{equation}
Recalling the definition of  Hermite-Gauss (HG) modes
\begin{equation}
HG_{1,0} = x f(r,z) \qquad
HG_{0,1} = y f(r,z),
\label{HG}
\end{equation}
and $x=r \cos(\phi)$, Eq.~\ref{eq:IR} describes a $HG_{1,0}$ mode rotated by $\varphi/2$ in the $(x, \, y)$ plane, except for a global phase factor that is irrelevant for our purpose.

\subsection{Qubit manipulation}

Manipulation of qubits as prepared above on the Bloch sphere at the experimenter's discretion can be performed using a set of two quadrupoles (QPs) as used in a MC. All (unitary) manipulations of a qubit correspond to a rotation on the Bloch sphere~\cite{U_v234_i_113456}.

In spherical coordinates, a general rotation by an angle $\theta$ around an axis given by the unit vector $\vec{n} = (n_X,n_Y,n_Z)^\top$ corresponds to the unitary operator
\begin{equation}
\op{R}_{\vec n}(\theta) = \ee^{-\ii \frac{\theta}{2} \vec{n} \cdot \vec{\sigma}} =
\cos\left(\frac{\theta}{2}\right) \mathbb{1} - \ii \sin\left(\frac{\theta}{2}\right) \vec{n} \cdot \vec{\sigma}.
\label{eq:Rn}
\end{equation}
where $\vec{\sigma}$ is the 3D vector of the Pauli matrices. 

The  MC performs a rotation of $\pi/2$ over the $X$ axis of the Bloch sphere shown in Fig.~\ref{fig:bloch-sphere}
\begin{multline}
\mat{R}_X (-\pi/2)=
\begin{pmatrix}
\cos(\pi/4) & -\ii\sin(\pi/4) \\
-\ii \sin(\pi/4) & \cos(\pi/4)	
\end{pmatrix} \\ =
\frac{1}{\sqrt 2}
\begin{pmatrix}
1 & -\ii \\
-\ii &  1	
\end{pmatrix} .
\label{eq:RX}
\end{multline}
Apart from a global phase, this is a $\rootnot$ quantum gate, usually abbreviated $\mat{RX}$. Note that the axes $X, Y, Z$ of the Bloch sphere in Fig.~\ref{fig:bloch-sphere}  must not be confused with the axes $x,y,z$ of the real space representation of the states, drawn in Fig.~\ref{fig:column}.

Applying the $\rootnot$ gate to the input qubit using Eqs.~\ref{eq:Iphi} and \ref{eq:RX} results in the output qubit
\begin{multline}
\ket{O_\varphi} \hat{=} \frac{1}{\sqrt{2}} \mat{RX} \begin{pmatrix} 1 \\ \ee^{\ii\varphi} \end{pmatrix} = 
\frac{1}{2} \begin{pmatrix} 1 - \ii \ee^{\ii \varphi} \\ -\ii + \ee^{\ii \varphi} \end{pmatrix} \\
= \ee^{\ii(\frac{\varphi}{2} - \frac{\pi}{4})} \begin{pmatrix} \cos(\frac{\varphi}{2} - \frac{\pi}{4}) \\ -\sin(\frac{\varphi}{2} - \frac{\pi}{4}) \end{pmatrix}
\label{eq:Ophi}
\end{multline}
It is readily apparent that for the eigenvectors of the transformation matrix, the trivial mapping occurs, namely
\begin{equation}
\begin{aligned}
\ket{I_{\SI{0}{\degree}}} = \ket{H} &\mapsto \ee^{-\ii \pi/4} \ket{H} \\
\ket{I_{\SI{180}{\degree}}} = \ket{V} &\mapsto \ee^{\ii \pi/4} \ket{V}.
\end{aligned}
\label{eq:trivial}
\end{equation}

\subsection{Qubit readout}

The third step is reading the output qubit. That means projecting it on the basis vectors of the Hilbert space, linked to a measurement device. Technically speaking, in the electron microscope these are pixels on a camera. This is a more subtle problem than it appears. As the intensity distribution of $\ket{R}$ and $\ket{L}$ in position space is identical (a ring), the two states cannot be distinguished and quantified by direct recording. 

Therefore, one of the OAM sorters proposed in the literature must be used, from early multi-pinhole interferometers~\cite{Clark2014} to holographic masks~\cite{Guzzinati2014} to the more recent OAM unwrappers~\cite{McMorran2017,Grillo2017,Pozzi2020,PRL_v126_i9_94802} which are based on a proposal for conformal mapping similar to light optics~\cite{Berkhout2010}. Their basis states --- in the present case eigenstates $\ket{R}, \, \ket{L}$ of the angular momentum operator $L_z$ --- become spatially separated in the sorter\footnote{Any readout basis can be selected by rotating the $L_z$ basis of the measurement device on the Bloch sphere into the readout basis. In principle, this can be achieved with a second MC, exactly as described for qubit manipulation.}.

\section{Experimental proof of principle}

\begin{figure}
	\centering\includegraphics[width=\columnwidth]{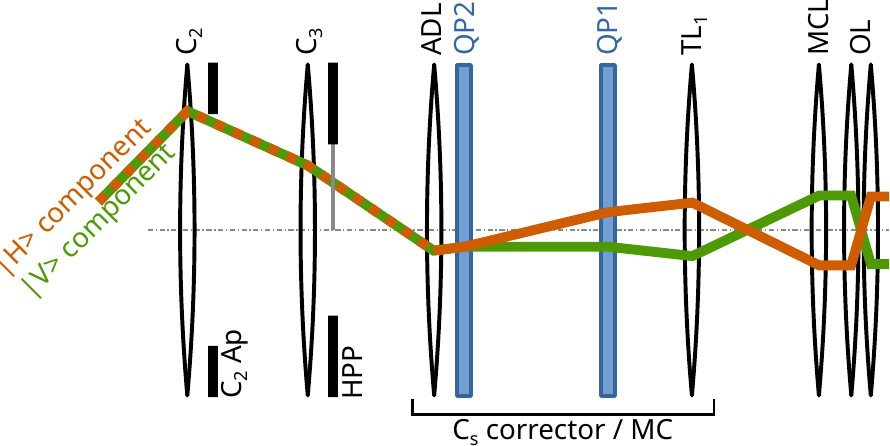}
	\caption{Sketch of the most important components of the probe-forming lens system, together with the geometric ray-paths for the $\ket{H}$ and $\ket{V}$ components.}
	\label{fig:raypath}
\end{figure}

We performed a proof-of-principle experiment on the J\"ulich PICO microscope, which is a monochromated double-C\textsubscript{s}-corrected (S)TEM instrument, together with numerical simulations of the beam propagation \cite{U_v204_i_p27} to analyze the beam shape and phase in experimentally inaccessible planes such as the MC entrance and exit planes. For this work, only the QPs in the probe-C\textsubscript{s} corrector were used to realize the MC~\cite{U_v204_i_p27,U_v229_i_113340,U_v234_i_113456}, i.e., the hexapoles were switched off. A simplified sketch of the general setup and the geometric ray-paths in the probe-forming lens system of the microscope is shown in Fig.~\ref{fig:raypath} \cite{U_v229_i_113340}. All experiments were carried out at \SI{200}{\kilo\electronvolt}.

\subsection{Qubit preparation}
Electrons closely resembling HG modes can be produced by several means, e.g., by exploiting the Aharanov-Bohm effect of a magnetic rod \cite{NP_v10_i1_p26}. Here, we used a Hilbert phase plate (HPP) \cite{U_v189_i_39,U_v229_i_113340} inserted in the C\textsubscript{3} aperture plane to phase-shift half the incident round beam by $\pi$. This resulted in a HG-like beam in the focal point, which is located in the input plane of the MC.

\begin{figure*}
	\centering\includegraphics[width=.75\textwidth]{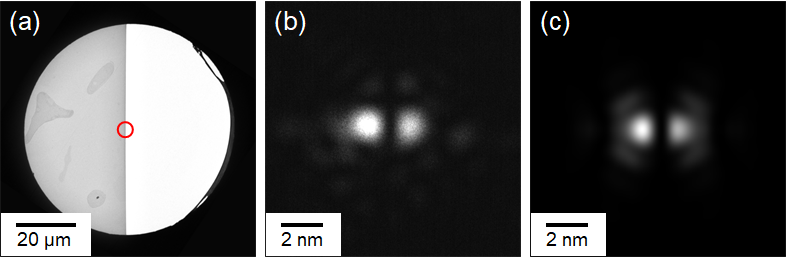}
	\caption{(a)~TEM image of the Hilbert phase-plate (HPP) with the red circle marking the actual illuminated area. (b)~Experimental image of the (unrotated) qubit with inactive quantum gate. (c)~Numerical simulation corresponding to (b).}
	\label{fig:hpp}
\end{figure*}

Fig.~\ref{fig:hpp}a shows a TEM image of the HPP. It consists of a conventional TEM aperture with a round hole with a diameter of \SI{70}{\micro\meter} half-covered by an electron-transparent phase-shifting layer system. The layer system consists of a \SI{11}{\nano\meter} metallic-glass Zirconium-Aluminium alloy (ZAC) covered by amorphous carbon (\SI{6}{\nano\meter} and \SI{12}{\nano\meter} thick, respectively) to prevent oxidization \cite{U_v229_i_113340}. The thicknesses were chosen to produce the required $\pi$ phase shift for the \SI{200}{\kilo\electronvolt} electrons.

Fig.~\ref{fig:hpp}b shows the intensity distribution of the beam in the MC output plane (i.e., the sample plane) for a disabled quantum gate (no quadrupole fields) to gauge the quality of the beam. It clearly is a nice, nano-meter-sized beam, albeit with slight intensity differences of the two lobes, steming from inelastic scattering in the HPP layer. The shape and size are in excellent agreement with the numerical simulations of the same setup, shown in Fig.~\ref{fig:hpp}c.

By rotating the MC coordinate system with respect to the HPP axis, arbitrary $\ket{I_\varphi}$ states can be prepared. In total, we performed four experiments. In our first experiment, we prepared a qubit 
\begin{equation}
\ket{I_{\SI{90}{\degree}}}  = \frac{1}{\sqrt{2}} ( \ket{R} + \ii \ket{L} ) = \ket{+}
\end{equation}
by rotating the $x$ axis of the MC by $\varphi/2 = \SI{45}{\degree}$ with respect to the HPP edge. According to Eq.~\ref{eq:IR} this is a $HG_{1,0}$ mode rotated by $\pi/4$. In the other experiments, we prepared the qubits
\begin{equation}
\begin{aligned}
\ket{I_{\SI{130}{\degree}}} &= \frac{1}{\sqrt{2}} \left( \ket{R} + \ee^{\ii \cdot \SI{130}{\degree}} \ket{L} \right) \\
\ket{I_{\SI{0}{\degree}}} &= \ket{H} \\
\ket{I_{\SI{180}{\degree}}} &= \ket{V}
\end{aligned}
\end{equation}
The last two acted as control experiments for the trivial mappings.

\subsection{Qubit manipulation}

Subsequently, the qubit was sent through the quantum gate realized by the MC. As stated above, the MC was realized by two QPs that were part of the (otherwise disabled) probe C\textsubscript{s} corrector. This set of QPs can perform arbitrary unitary transformations on the Bloch sphere by combining two types of operations~\cite{U_v234_i_113456}: (i) a rotation around the Bloch sphere's $Z$ axis can be performed by a rotation of the QPs' coordinate system w.r.t. the input plane, i.e., changing the field axes by changing the current through the magnetic coils\footnote{This requires pairs of QPs, though}; (ii) a rotation around the Bloch sphere's $Y$ axis can be performed by a relative phase shift between the $\ket{H}$ and $\ket{V}$ components (in the QPs' frame of reference) as one of them goes through a focal point (see ref.~\ref{fig:raypath}).

Given the geometry of the microscope, the required focal lengths can be calculated as detailed in~\cite{U_v234_i_113456}. Moreover, the relationship between electric current and focal length can be calibrated for each lens by focusing in specific, fixed planes (e.g., focusing the beam in the sample plane, sharply imaging the various aperture planes, etc.). Once the calibration curves are known, the required focal lengths can easily be set. Note that both the calibration curves and the required focal lengths will vary from instrument to instrument due to slight variations and tolerances in manufacturing.

In our case, we set up the MC such that it acted as an $\mat{RX}$ gate. For the four experiments, this resulted in the mappings
\begin{equation}
\begin{aligned}
	\ket{I_{\SI{90}{\degree}}} &\mapsto \ket{O_{\SI{90}{\degree}}} = \ket{R} \\
	\ket{I_{\SI{130}{\degree}}} &\mapsto \ket{O_{\SI{130}{\degree}}} = \ee^{\ii \cdot \SI{20}{\degree}} \left( 0.9397 \ket{R} -0.3420 \ket{L} \right) \\
	\ket{I_{\SI{0}{\degree}}} &\mapsto \ket{O_{\SI{0}{\degree}}} = \ee^{-\ii \pi/4}\ket{H} \\
	\ket{I_{\SI{180}{\degree}}} &\mapsto \ket{O_{\SI{180}{\degree}}} = \ee^{\ii \pi/4}\ket{V}
\end{aligned}
\end{equation}
as given in Eq.~\ref{eq:Ophi}.

\begin{figure}
	\centering\includegraphics[width=\columnwidth]{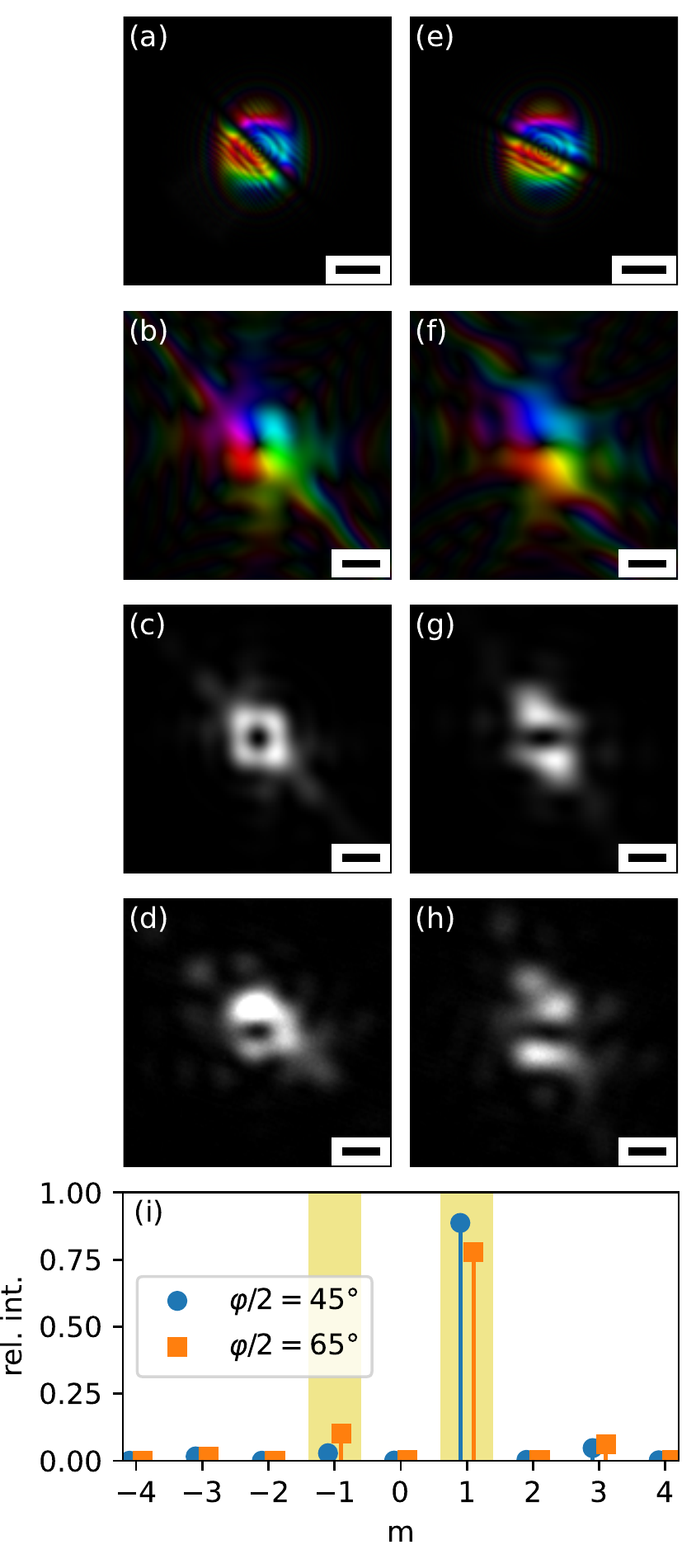}
	\caption{(a) Simulated input qubit for $\varphi/2 = \SI{45}{\degree}$. (b) Simulated output qubit after sending the input qubit (a) through the gate. (c) Intensity distribution of (b). (d) Experimentally observed intensity corresponding to (c). (e--h) Analogous data for $\varphi/2 = \SI{65}{\degree}$. (i) Intensity histogram for different OAM eigenvalues for the two output qubits. The scale bars in (a, e) denote \SI{500}{\nano\meter}, all other scale bars denote \SI{2}{\nano\meter}.}
	\label{fig:exp2}
\end{figure}

\subsection{Qubit readout}

Subsequently, the beam was sent through the objective lens and the projection system and was finally observed on a CCD.

Fig.~\ref{fig:exp2} shows a comparison between the experimental beam in the sample plane and the corresponding simulation of propagation through the column. OAM analysis of the output qubits was done numerically\footnote{OAM sorters have been tested successfully elsewhere~\cite{McMorran2017,Grillo2017,Pozzi2020,PRL_v126_i9_94802}. Since the implementation of a sorter in the J{\"u}lich PICO microscope is currently not feasible we chose a numerical approach (see appendix~\ref{sec:oam-calc}).}.
The results are shown as histogram in Fig.~\ref{fig:exp2}. It is clearly visible that (apart from some impurities due to HPP imperfections \cite{U_v229_i_113340,U_v204_i_p27}) the output qubit in the $\varphi/2 = \SI{45}{\degree}$ case consists essentially of the $m=1$ component (corresponding to $\ket{R}$), whereas in the $\varphi/2 = \SI{65}{\degree}$ case, the output qubit features roughly $(-0.342)^2 \approx 0.117$ relative intensity of the $m=-1$ component (corresponding to $\ket{L}$) over the impurity background.

\begin{figure}
	\centering\includegraphics[width=\columnwidth]{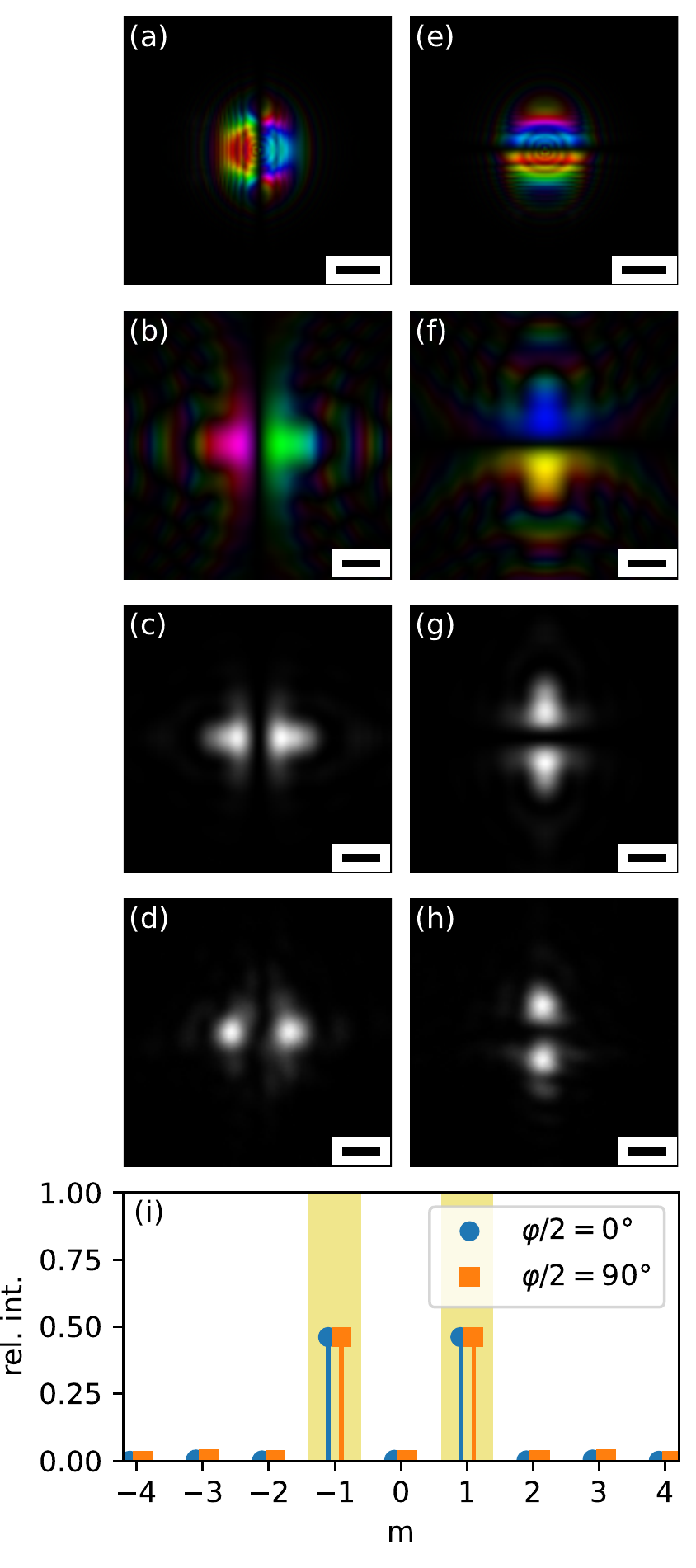}
	\caption{(a) Simulation of the input qubit corresponding to an $\ket{H}$ state produced by the HPP. (b) Output qubit corresponding to (a). (c) Intensity distribution of the output qubit. (d) Experimental intensity distribution. (e--h) Analogous data for an input $\ket{V}$ state. (i) Histogram of the relative intensities of the different vortex components for the output qubits. The scale bars in (a, e) denote \SI{500}{\nano\meter}, all other scale bars denote \SI{2}{\nano\meter}.}
	\label{fig:trivial-mapping}
\end{figure}

Likewise, Fig.~\ref{fig:trivial-mapping} shows the experimental and simulated data for the control experiments with $\ket{I_{\SI{0}{\degree}}} = \ket{H}$ and $\ket{I_{\SI{180}{\degree}}} = \ket{V}$. It is clearly evident that in both cases, the typical two-lobed structure as well as the orientation is preserved upon passage through the quantum gate, as expected. As any linear operator on a 2D Hilbert space is uniquely defined by its action on two linearly independent (basis) vectors, these results further support the conclusion that the MC acts as a general $\rootnot$ gate.

\section{Discussion and Conclusion}

High-end TEMs --- instruments of utmost stability,  spatial and energy resolution, sophisticated lens systems, ultra sensitive detectors and pulsed electron sources with repetition rates of the order of \si{\mega\hertz} --- provide an ideal scenario to extend qubit manipulation from photons, superconducting circuits or ions to freely floating electrons. This novel platform for the study of qubits has several genuine features: qubits can be tailored from \si{\nano\meter} to \si{\micro\meter} size; in free space, they are topologically protected~\cite{PRA_v87_i_p33834,Kitaev2003}; there is no need for cryogenic temperatures; they reveal high decoherence times and essentially no relaxation because the energies of the basis states are identical. Recent work on entanglement in electron microscopy~\cite{Okamoto2014,Schatt2018,JoESaRP_v241_p146810,Haindl2022,Meier2022,U_v241_i_113594,SA_v8_i47_eabo7853} could provide opportunities for 2-qubit gates in non-separable systems. 

It is not very likely that free electron qubits will enter the race to quantum computing applications, although deceleration in electrostatic fields could well increase their presently short lifetimes. Their most attractive aspect is perhaps a broad range of tunable perturbations of the  qubits via controlled interaction with electromagnetic radiation or matter on their way down the microscope column, in order to study the robustness of quantum gate operations, their fidelity and reliability \cite{PRL_v126_i23_233403}. 

\section*{Acknowledgements}
PS thanks D. Karlovets and P. Hommelhof for valuable comments. SL, TS, and PS gratefully acknowledge financial support by the Austrian Science Fund under the projects P29687-N36 and I4309-N36. CEOS GmbH has received funding from the European Union's Horizon 2020 research and innovation programme under grant agreement No. 823717-ESTEEM3. PHL and REDB acknowledge funding from European Union’s Horizon 2020 research and innovation programme under grant agreement No. 856538 (``3D MAGiC''). MO, MD and DG acknowledge funding of this work by the German Research Foundation (Deutsche Forschungsgemeinschaft) under contract Ge~841/26. The authors acknowledge TU Wien Bibliothek for financial support through its Open Access Funding Programme.

\appendix

\section{Numerical Calculation of the OAM spectrum}
\label{sec:oam-calc}

As the OAM operator $\op{L}_z$ commutes with the (spherically symmetric) free-space Hamiltonian, it is possible to find a simultaneous eigenbasis $\{ \ket{m} \}$ for the two operators. As $\op{L}_z \ket{m} = \hbar m \ket{m}$, the real-space angular dependence of those states is given by
\[
	\braket{\vec{r} | m} = c(r, z) \ee^{\ii m \phi}.
\]
With this, all free-space wavefunctions $\ket{\psi}$ can be written in real-space representation as
\[
	\braket{\vec{r} | \psi} = \sum_m c_m(r, z) \ee^{\ii m \phi}
\]
and the wave-function's OAM spectrum $I(m)$ is given by
\begin{equation}
	I(m) = \iint |c_m(r, z)|^2 r \dd r \dd z.
\end{equation}
Theoretically, these integrals necessarily converge for all $m$ for any normalized $\psi$. In practice, the integrals are typically evaluated within a bounded region where the wavefunction is non-negligible (and above noise level).

To actually calculate the OAM spectrum, we first transformed the wavefunction to a polar representation $(r, \phi, z)$, using bilinear interpolation where necessary. Then, we Fourier-transformed with respect to the $\phi$ component~\cite{ACSA_v75_p902}, resulting in
\begin{align*}
\tilde{\psi}(r, l, z)
&= \frac{1}{2\pi}\int_{0}^{2\pi} \psi(r, \phi, z) \ee^{-\ii l \phi} \dd \phi \\
&= \frac{1}{2\pi}\sum_m c_m(r, z) \int_{0}^{2\pi} \ee^{-\ii l \phi} \ee^{\ii m \phi} \dd \phi \\
&= \frac{1}{2\pi}\sum_m c_m(r, z) \cdot 2 \pi \delta_{l,m} \\
&= c_l(r, z).
\end{align*}
Finally, the spectrum is given simply by the norm
\[
	I_m = \iint |\tilde{\psi}(r, m, z)|^2 r \dd r \dd z.
\]


\begin{thebibliography}{53}
\providecommand{\natexlab}[1]{#1}
\providecommand{\url}[1]{\texttt{#1}}
\expandafter\ifx\csname urlstyle\endcsname\relax
  \providecommand{\doi}[1]{doi: #1}\else
  \providecommand{\doi}{doi: \begingroup \urlstyle{rm}\Url}\fi

\bibitem[Rotunno et~al.(2019)Rotunno, Tavabi, Yucelen, Frabboni, Borkowski,
  Karimi, McMorran, and Grillo]{PRA_v11_i_p44072}
E.~Rotunno, A.H. Tavabi, E.~Yucelen, S.~Frabboni, R.E.~Dunin Borkowski,
  E.~Karimi, B.J. McMorran, and V.~Grillo.
\newblock Electron-beam shaping in the transmission electron microscope:
  Control of electron-beam propagation along atomic columns.
\newblock \emph{Phys. Rev. Appl.}, 11\penalty0 (4):\penalty0 044072, April
  2019.
\newblock \doi{10.1103/physrevapplied.11.044072}.

\bibitem[Hammer et~al.(2015)Hammer, Thomas, Weber, and
  Hommelhoff]{Hommelhof2015}
J.~Hammer, S.~Thomas, P.~Weber, and P.~Hommelhoff.
\newblock Microwave chip-based beam splitter for low-energy guided electrons.
\newblock \emph{Phys. Rev. Lett.}, 114\penalty0 (25):\penalty0 254801, 2015.
\newblock \doi{10.1103/PhysRevLett.114.254801}.

\bibitem[Schachinger et~al.(2017)Schachinger, Löffler, Steiger-Thirsfeld,
  Stöger-Pollach, Schneider, Pohl, Rellinghaus, and
  Schattschneider]{Schachinger2017}
T.~Schachinger, S.~Löffler, A.~Steiger-Thirsfeld, M.~Stöger-Pollach,
  S.~Schneider, D.~Pohl, B.~Rellinghaus, and P.~Schattschneider.
\newblock {EMCD} with an electron vortex filter: Limitations and possibilities.
\newblock \emph{Ultramicroscopy}, 179:\penalty0 15--23, 2017.
\newblock \doi{10.1016/j.ultramic.2017.03.019}.

\bibitem[Verbeeck et~al.(2013)Verbeeck, Tian, and
  Van~Tendeloo]{VerbeeckAdvMat2013}
J.~Verbeeck, H.~Tian, and G.~Van~Tendeloo.
\newblock How to manipulate nanoparticles with an electron beam?
\newblock \emph{Adv. Mater.}, 25\penalty0 (8):\penalty0 1114--1117, 2013.
\newblock \doi{10.1002/adma.201204206}.

\bibitem[Franke-Arnold et~al.(2008)Franke-Arnold, Allen, and
  Padgett]{Franke2008}
S.~Franke-Arnold, L.~Allen, and M.~Padgett.
\newblock Advances in optical angular momentum.
\newblock \emph{Laser Photonics Rev.}, 2\penalty0 (4):\penalty0 299--313, 2008.
\newblock \doi{10.1002/lpor.200810007}.

\bibitem[Babazadeh et~al.(2017)Babazadeh, Erhard, Wang, Malik, Nouroozi, Krenn,
  and Zeilinger]{Zeilinger2017}
A.~Babazadeh, M.~Erhard, F.~Wang, M.~Malik, R.~Nouroozi, M.~Krenn, and
  A.~Zeilinger.
\newblock High-dimensional single-photon quantum gates: Concepts and
  experiments.
\newblock \emph{Phys. Rev. Lett.}, 119:\penalty0 180510, Nov 2017.
\newblock \doi{10.1103/PhysRevLett.119.180510}.

\bibitem[Juchtmans et~al.(2015)Juchtmans, B\'ech\'e, Abakumov, Batuk, and
  Verbeeck]{Juchtmans2015}
R.~Juchtmans, A.~B\'ech\'e, A.~Abakumov, M.~Batuk, and J.~Verbeeck.
\newblock Using electron vortex beams to determine chirality of crystals in
  transmission electron microscopy.
\newblock \emph{Phys. Rev. B}, 91:\penalty0 094112, Mar 2015.
\newblock \doi{10.1103/PhysRevB.91.094112}.

\bibitem[Vanacore et~al.(2018)Vanacore, Madan, Berruto, Wang, Pomarico, Lamb,
  McGrouther, Kaminer, Barwick, Garcia De~Abajo, and Carbone]{Carbone2018}
G.~M. Vanacore, I.~Madan, G.~Berruto, K.~Wang, E.~Pomarico, R.~J. Lamb,
  D.~McGrouther, I.~Kaminer, B.~Barwick, F.~J. Garcia De~Abajo, and F.~Carbone.
\newblock Attosecond coherent control of free-electron wave functions using
  semi-infinite light fields.
\newblock \emph{Nat. Commun.}, 9\penalty0 (1):\penalty0 2694, 2018.
\newblock \doi{10.1038/s41467-018-05021-x}.

\bibitem[Feist et~al.(2015)Feist, Echternkamp, Schauss, Yalunin, Sch{\"a}fer,
  and Ropers]{Feist2015200}
A.~Feist, K.E. Echternkamp, J.~Schauss, S.V. Yalunin, S.~Sch{\"a}fer, and
  C.~Ropers.
\newblock Quantum coherent optical phase modulation in an ultrafast
  transmission electron microscope.
\newblock \emph{Nature}, 521\penalty0 (7551):\penalty0 200--203, 2015.
\newblock \doi{10.1038/nature14463}.

\bibitem[Kealhofer et~al.(2016)Kealhofer, Schneider, Ehberger, Ryabov, Krausz,
  and Baum]{Baum2016}
C.~Kealhofer, W.~Schneider, D.~Ehberger, A.~Ryabov, F.~Krausz, and P.~Baum.
\newblock All-optical control and metrology of electron pulses.
\newblock \emph{Science}, 352\penalty0 (6284):\penalty0 429--433, 2016.
\newblock \doi{10.1126/science.aae0003}.

\bibitem[Sch\"onenberger et~al.(2019)Sch\"onenberger, Mittelbach, Yousefi,
  McNeur, Niedermayer, and Hommelhoff]{Hommelhoff2019}
N.~Sch\"onenberger, A.~Mittelbach, P.~Yousefi, J.~McNeur, U.~Niedermayer, and
  P.~Hommelhoff.
\newblock Generation and characterization of attosecond microbunched electron
  pulse trains via dielectric laser acceleration.
\newblock \emph{Phys. Rev. Lett.}, 123\penalty0 (26):\penalty0 264803, 2019.
\newblock \doi{10.1103/PhysRevLett.123.264803}.

\bibitem[Bliokh et~al.(2007)Bliokh, Bliokh, Savel'ev, and Nori]{Bliokh2007}
K.~Y. Bliokh, Y.~P. Bliokh, S.~Savel'ev, and F.~Nori.
\newblock Semiclassical dynamics of electron wave packet states with phase
  vortices.
\newblock \emph{Phys. Rev. Lett.}, 99\penalty0 (19), 2007.
\newblock \doi{10.1103/PhysRevLett.99.190404}.

\bibitem[Bliokh et~al.(2011)Bliokh, Dennis, and Nori]{Bliokh2011}
K.~Y. Bliokh, M.~R. Dennis, and F.~Nori.
\newblock Relativistic electron vortex beams: Angular momentum and spin-orbit
  interaction.
\newblock \emph{Phys. Rev. Lett.}, 107\penalty0 (17), 2011.
\newblock \doi{10.1103/PhysRevLett.107.174802}.

\bibitem[Verbeeck et~al.(2010)Verbeeck, Tian, and
  Schattschneider]{Verbeeck2010}
J.~Verbeeck, H.~Tian, and P.~Schattschneider.
\newblock Production and application of electron vortex beams.
\newblock \emph{Nature}, 467\penalty0 (7313):\penalty0 301--304, 2010.
\newblock \doi{10.1038/nature09366}.

\bibitem[Uchida and Tonomura(2010)]{N_v464_i_p737}
M.~Uchida and A.~Tonomura.
\newblock Generation of electron beams carrying orbital angular momentum.
\newblock \emph{Nat.}, 464:\penalty0 737--739, 04 2010.
\newblock \doi{10.1038/nature08904}.

\bibitem[Bliokh et~al.(2012)Bliokh, Schattschneider, Verbeeck, and
  Nori]{Bliokh2012}
K.Y. Bliokh, P.~Schattschneider, J.~Verbeeck, and F.~Nori.
\newblock Electron vortex beams in a magnetic field: A new twist on {L}andau
  levels and {A}haronov-{B}ohm states.
\newblock \emph{Phys. Rev. X}, 2\penalty0 (4):\penalty0 041011, 2012.
\newblock \doi{10.1103/PhysRevX.2.041011}.

\bibitem[Schattschneider et~al.(2014)Schattschneider, Schachinger,
  Stöger-Pollach, Löffler, Steiger-Thirsfeld, Bliokh, and
  Nori]{NC_v5_i_p4586}
P.~Schattschneider, T.~Schachinger, M.~Stöger-Pollach, S.~Löffler,
  A.~Steiger-Thirsfeld, K.~Y. Bliokh, and F.~Nori.
\newblock {Imaging the dynamics of free-electron Landau states}.
\newblock \emph{Nat. Commun.}, 5:\penalty0 4586, August 2014.
\newblock \doi{10.1038/ncomms5586}.

\bibitem[Guzzinati et~al.(2013)Guzzinati, Schattschneider, Bliokh, Nori, and
  Verbeeck]{PRL_v110_i_p93601}
G.~Guzzinati, P.~Schattschneider, K.~Y. Bliokh, F.~Nori, and J.~Verbeeck.
\newblock {Observation of the Larmor and Gouy Rotations with Electron Vortex
  Beams}.
\newblock \emph{Phys. Rev. Lett.}, 110:\penalty0 093601, February 2013.
\newblock \doi{10.1103/PhysRevLett.110.093601}.

\bibitem[Schachinger et~al.(2015)Schachinger, Löffler, Stöger-Pollach, and
  Schattschneider]{U_v158_i_p17}
T.~Schachinger, S.~Löffler, M.~Stöger-Pollach, and P.~Schattschneider.
\newblock Peculiar rotation of electron vortex beams.
\newblock \emph{Ultramicroscopy}, 158:\penalty0 17--25, November 2015.
\newblock ISSN 0304-3991.
\newblock \doi{10.1016/j.ultramic.2015.06.004}.

\bibitem[Bliokh et~al.(2017)Bliokh, Ivanov, Guzzinati, Clark, Van~Boxem,
  Béché, Juchtmans, Alonso, Schattschneider, Nori, and Verbeeck]{Bliokh2017}
K.~Y. Bliokh, I.~P. Ivanov, G.~Guzzinati, L.~Clark, R.~Van~Boxem, A.~Béché,
  R.~Juchtmans, M.~A. Alonso, P.~Schattschneider, F.~Nori, and J.~Verbeeck.
\newblock Theory and applications of free-electron vortex states.
\newblock \emph{Phys. Rep.}, 690:\penalty0 1--70, 2017.
\newblock \doi{10.1016/j.physrep.2017.05.006}.

\bibitem[Larsen et~al.(2019)Larsen, Guo, Breum, Neergaard-Nielsen, and
  Andersen]{qubits-Larsen}
M.~V. Larsen, X.~Guo, C.~R. Breum, J.~S. Neergaard-Nielsen, and U.~L. Andersen.
\newblock Deterministic generation of a two-dimensional cluster state.
\newblock \emph{Science}, 366\penalty0 (6463):\penalty0 369--372, 2019.
\newblock \doi{10.1126/science.aay4354}.

\bibitem[Brown et~al.(2021)Brown, Chiaverini, Sage, and Häffner]{qubits-Brown}
K.~R. Brown, J.~Chiaverini, J.~M. Sage, and H.~Häffner.
\newblock Materials challenges for trapped-ion quantum computers.
\newblock \emph{Nat. Rev. Mater.}, 6\penalty0 (10):\penalty0 892--905, 2021.
\newblock \doi{10.1038/s41578-021-00292-1}.

\bibitem[Kjaergaard et~al.(2020)Kjaergaard, Schwartz, Braumüller, Krantz,
  Wang, Gustavsson, and Oliver]{qubits-Kjaergaard}
M.~Kjaergaard, M.~E. Schwartz, J.~Braumüller, P.~Krantz, J.~I.~. Wang,
  S.~Gustavsson, and W.~D. Oliver.
\newblock Superconducting qubits: Current state of play.
\newblock \emph{Annu. Rev. Conden. Ma. P.}, 11:\penalty0 369--395, 2020.
\newblock \doi{10.1146/annurev-conmatphys-031119-050605}.

\bibitem[Bradley et~al.(2019)Bradley, Randall, Abobeih, Berrevoets, Degen,
  Bakker, Markham, Twitchen, and Taminiau]{qubits-Bradley}
C.~E. Bradley, J.~Randall, M.~H. Abobeih, R.~C. Berrevoets, M.~J. Degen, M.~A.
  Bakker, M.~Markham, D.~J. Twitchen, and T.~H. Taminiau.
\newblock A ten-qubit solid-state spin register with quantum memory up to one
  minute.
\newblock \emph{Phys. Rev. X}, 9\penalty0 (3), 2019.
\newblock \doi{10.1103/PhysRevX.9.031045}.

\bibitem[Buluta et~al.(2011)Buluta, Ashhab, and Nori]{RoPiP_v74_i10_104401}
I.~Buluta, S.~Ashhab, and F.~Nori.
\newblock Natural and artificial atoms for quantum computation.
\newblock \emph{Rep. Prog. Phys.}, 74\penalty0 (10):\penalty0 104401, sep 2011.
\newblock \doi{10.1088/0034-4885/74/10/104401}.

\bibitem[Chatterjee et~al.(2021)Chatterjee, Stevenson, De~Franceschi, Morello,
  de~Leon, and Kuemmeth]{qubits-Chatterjee}
A.~Chatterjee, P.~Stevenson, S.~De~Franceschi, A.~Morello, N.~P. de~Leon, and
  F.~Kuemmeth.
\newblock Semiconductor qubits in practice.
\newblock \emph{Nature Reviews Physics}, 3\penalty0 (3):\penalty0 157--177,
  2021.
\newblock \doi{10.1038/s42254-021-00283-9}.
\newblock Cited By :91.

\bibitem[Reinhardt et~al.(2021)Reinhardt, Mechel, Lynch, and
  Kaminer]{Reinhardt2021}
O.~Reinhardt, C.~Mechel, M.~Lynch, and I.~Kaminer.
\newblock Free-electron qubits.
\newblock \emph{Ann. Phys.}, 533\penalty0 (2):\penalty0 2000254, 2021.
\newblock \doi{10.1002/andp.202000254}.

\bibitem[Ruimy et~al.(2021)Ruimy, Gorlach, Mechel, Rivera, and
  Kaminer]{PRL_v126_i23_233403}
R.~Ruimy, A.~Gorlach, C.~Mechel, N.~Rivera, and I.~Kaminer.
\newblock Toward atomic-resolution quantum measurements with coherently shaped
  free electrons.
\newblock \emph{Phys. Rev. Lett.}, 126\penalty0 (23):\penalty0 233403, jun
  2021.
\newblock \doi{10.1103/physrevlett.126.233403}.

\bibitem[Tsarev et~al.(2021)Tsarev, Ryabov, and Baum]{PRR_v3_i4_43033}
M.~V. Tsarev, A.~Ryabov, and P.~Baum.
\newblock Free-electron qubits and maximum-contrast attosecond pulses via
  temporal talbot revivals.
\newblock \emph{Phys. Rev. Research}, 3\penalty0 (4):\penalty0 043033, oct
  2021.
\newblock \doi{10.1103/physrevresearch.3.043033}.

\bibitem[Löffler(2022)]{U_v234_i_113456}
S.~Löffler.
\newblock Unitary two-state quantum operators realized by quadrupole fields in
  the electron microscope.
\newblock \emph{Ultramicroscopy}, 234:\penalty0 113456, 2022.
\newblock \doi{10.1016/j.ultramic.2021.113456}.

\bibitem[Schattschneider et~al.(2012)Schattschneider, St\"oger-Pollach, and
  Verbeeck]{Schattschneider2012}
P.~Schattschneider, M.~St\"oger-Pollach, and J.~Verbeeck.
\newblock Novel vortex generator and mode converter for electron beams.
\newblock \emph{Phys. Rev. Lett.}, 109\penalty0 (8):\penalty0 084801, 2012.
\newblock \doi{10.1103/PhysRevLett.109.084801}.

\bibitem[Schachinger et~al.(2021)Schachinger, Hartel, Lu, Löffler, Obermair,
  Dries, Gerthsen, Dunin-Borkowski, and Schattschneider]{U_v229_i_113340}
T.~Schachinger, P.~Hartel, P.~Lu, S.~Löffler, M.~Obermair, M.~Dries,
  D.~Gerthsen, R.~E. Dunin-Borkowski, and P.~Schattschneider.
\newblock Experimental realisation of a $\pi/2$ vortex mode converter for
  electrons using a spherical aberration corrector.
\newblock \emph{Ultramicroscopy}, 229:\penalty0 113340, 2021.
\newblock \doi{10.1016/j.ultramic.2021.113340}.

\bibitem[Karlovets(2018)]{Karlovets2018}
D.~Karlovets.
\newblock Relativistic vortex electrons: Paraxial versus nonparaxial regimes.
\newblock \emph{Phys. Rev. A}, 98:\penalty0 012137, Jul 2018.
\newblock \doi{10.1103/PhysRevA.98.012137}.

\bibitem[Clark et~al.(2014)Clark, Béché, Guzzinati, and Verbeeck]{Clark2014}
L.~Clark, A.~Béché, G.~Guzzinati, and J.~Verbeeck.
\newblock Quantitative measurement of orbital angular momentum in electron
  microscopy.
\newblock \emph{Physical Review A - Atomic, Molecular, and Optical Physics},
  89\penalty0 (5):\penalty0 053818, 2014.
\newblock \doi{10.1103/PhysRevA.89.053818}.

\bibitem[Guzzinati et~al.(2014)Guzzinati, Clark, Béché, and
  Verbeeck]{Guzzinati2014}
G.~Guzzinati, L.~Clark, A.~Béché, and J.~Verbeeck.
\newblock Measuring the orbital angular momentum of electron beams.
\newblock \emph{Physical Review A - Atomic, Molecular, and Optical Physics},
  89\penalty0 (2):\penalty0 025803, 2014.
\newblock \doi{10.1103/PhysRevA.89.025803}.

\bibitem[McMorran et~al.(2017)McMorran, Harvey, and Lavery]{McMorran2017}
B.~J. McMorran, T.~R. Harvey, and M.~P.~J. Lavery.
\newblock Efficient sorting of free electron orbital angular momentum.
\newblock \emph{New J. Phys.}, 19\penalty0 (2):\penalty0 023053, 2017.
\newblock \doi{10.1088/1367-2630/aa5f6f}.

\bibitem[Grillo et~al.(2017)Grillo, Tavabi, Venturi, Larocque, Balboni,
  Gazzadi, Frabboni, Lu, Mafakheri, Bouchard, Dunin-Borkowski, Boyd, Lavery,
  Padgett, and Karimi]{Grillo2017}
V.~Grillo, A.~H. Tavabi, F.~Venturi, H.~Larocque, R.~Balboni, G.~C. Gazzadi,
  S.~Frabboni, P.~. Lu, E.~Mafakheri, F.~Bouchard, R.~E. Dunin-Borkowski, R.~W.
  Boyd, M.~P.~J. Lavery, M.~J. Padgett, and E.~Karimi.
\newblock Measuring the orbital angular momentum spectrum of an electron beam.
\newblock \emph{Nat. Commun.}, 8:\penalty0 15536, 2017.
\newblock \doi{10.1038/ncomms15536}.

\bibitem[Pozzi et~al.(2020)Pozzi, Grillo, Lu, Tavabi, Karimi, and
  Dunin-Borkowski]{Pozzi2020}
G.~Pozzi, V.~Grillo, P.~Lu, A.~H. Tavabi, E.~Karimi, and R.~E. Dunin-Borkowski.
\newblock Design of electrostatic phase elements for sorting the orbital
  angular momentum of electrons.
\newblock \emph{Ultramicroscopy}, 208:\penalty0 112861, 2020.
\newblock \doi{10.1016/j.ultramic.2019.112861}.

\bibitem[Tavabi et~al.(2021)Tavabi, Rosi, Rotunno, Roncaglia, Belsito,
  Frabboni, Pozzi, Gazzadi, Lu, Nijland, Ghosh, Tiemeijer, Karimi,
  Dunin-Borkowski, and Grillo]{PRL_v126_i9_94802}
A.~H. Tavabi, P.~Rosi, E.~Rotunno, A.~Roncaglia, L.~Belsito, S.~Frabboni,
  G.~Pozzi, G.~C. Gazzadi, P.~Lu, R.~Nijland, M.~Ghosh, P.~Tiemeijer,
  E.~Karimi, R.~E. Dunin-Borkowski, and V.~Grillo.
\newblock Experimental demonstration of an electrostatic orbital angular
  momentum sorter for electron beams.
\newblock \emph{Phys. Rev. Lett.}, 126\penalty0 (9):\penalty0 094802, mar 2021.
\newblock \doi{10.1103/physrevlett.126.094802}.

\bibitem[Berkhout et~al.(2010)Berkhout, Lavery, Courtial, Beijersbergen, and
  Padgett]{Berkhout2010}
G.~C.~G. Berkhout, M.~P.~J. Lavery, J.~Courtial, M.~W. Beijersbergen, and M.~J.
  Padgett.
\newblock Efficient sorting of orbital angular momentum states of light.
\newblock \emph{Phys. Rev. Lett.}, 105\penalty0 (15):\penalty0 153601, 2010.
\newblock \doi{10.1103/PhysRevLett.105.153601}.

\bibitem[Kramberger et~al.(2019)Kramberger, Löffler, Schachinger, Hartel,
  Zach, and Schattschneider]{U_v204_i_p27}
C.~Kramberger, S.~Löffler, T.~Schachinger, P.~Hartel, J.~Zach, and
  P.~Schattschneider.
\newblock \texorpdfstring{$\pi$}{π}/2 mode converters and vortex generators
  for electrons.
\newblock \emph{Ultramicroscopy}, 204:\penalty0 27--33, September 2019.
\newblock \doi{10.1016/j.ultramic.2019.05.003}.

\bibitem[Béché et~al.(2013)Béché, Van~Boxem, Van~Tendeloo, and
  Verbeeck]{NP_v10_i1_p26}
A.~Béché, R.~Van~Boxem, G.~Van~Tendeloo, and J.~Verbeeck.
\newblock Magnetic monopole field exposed by electrons.
\newblock \emph{Nat. Phys.}, 10\penalty0 (1):\penalty0 26–29, December 2013.
\newblock ISSN 1745-2481.
\newblock \doi{10.1038/nphys2816}.

\bibitem[Dries et~al.(2018)Dries, Obermair, Hettler, Hermann, Seemann,
  Seifried, Ulrich, Fischer, and Gerthsen]{U_v189_i_39}
M.~Dries, M.~Obermair, S.~Hettler, P.~Hermann, K.~Seemann, F.~Seifried,
  S.~Ulrich, R.~Fischer, and D.~Gerthsen.
\newblock Oxide-free {aC}/\ce{Zr_{0.65}Al_{0.075}Cu_{0.275}}/{aC} phase plates
  for transmission electron microscopy.
\newblock \emph{Ultramicroscopy}, 189:\penalty0 39--45, jun 2018.
\newblock \doi{10.1016/j.ultramic.2018.03.003}.

\bibitem[Lubk et~al.(2013)Lubk, Clark, Guzzinati, and
  Verbeeck]{PRA_v87_i_p33834}
A.~Lubk, L.~Clark, G.~Guzzinati, and J.~Verbeeck.
\newblock Topological analysis of paraxially scattered electron vortex beams.
\newblock \emph{Phys. Rev. A}, 87:\penalty0 033834, March 2013.
\newblock \doi{10.1103/PhysRevA.87.033834}.

\bibitem[Kitaev(2003)]{Kitaev2003}
A.~Y. Kitaev.
\newblock Fault-tolerant computation by anyons.
\newblock \emph{Ann. Phys.}, 303:\penalty0 2--30, 2003.
\newblock \doi{10.1016/S0003-4916(02)00018-0}.

\bibitem[Okamoto(2014)]{Okamoto2014}
H.~Okamoto.
\newblock Measurement errors in entanglement-assisted electron microscopy.
\newblock \emph{Physical Review A - Atomic, Molecular, and Optical Physics},
  89\penalty0 (6):\penalty0 063828, 2014.
\newblock \doi{10.1103/PhysRevA.89.063828}.

\bibitem[Schattschneider and L{\"o}ffler(2018)]{Schatt2018}
P.~Schattschneider and S.~L{\"o}ffler.
\newblock Entanglement and decoherence in electron microscopy.
\newblock \emph{Ultramicroscopy}, 190:\penalty0 39--44, 2018.
\newblock \doi{10.1016/j.ultramic.2018.04.007}.

\bibitem[Schattschneider et~al.(2020)Schattschneider, Löffler, Gollisch, and
  Feder]{JoESaRP_v241_p146810}
P.~Schattschneider, S.~Löffler, H.~Gollisch, and R.~Feder.
\newblock Entanglement and entropy in electron--electron scattering.
\newblock \emph{J. Electron Spectrosc. Relat. Phenom.}, 241:\penalty0 146810,
  2020.
\newblock \doi{10.1016/j.elspec.2018.11.009}.

\bibitem[Haindl et~al.(2023)Haindl, Feist, Domröse, Möller, Gaida, Yalunin,
  and Ropers]{Haindl2022}
R.~Haindl, A.~Feist, T.~Domröse, M.~Möller, J.~H. Gaida, S.~V. Yalunin, and
  C.~Ropers.
\newblock Coulomb-correlated electron number states in a transmission electron
  microscope beam.
\newblock \emph{Nature Physics}, 2023.
\newblock \doi{10.1038/s41567-023-02067-7}.

\bibitem[Meier et~al.(2023)Meier, Heimerl, and Hommelhoff]{Meier2022}
S.~Meier, J.~Heimerl, and P.~Hommelhoff.
\newblock Few-electron correlations after ultrafast photoemission from
  nanometric needle tips.
\newblock \emph{Nature Physics}, 2023.
\newblock \doi{10.1038/s41567-023-02059-7}.

\bibitem[Scheucher et~al.(2022)Scheucher, Schachinger, Spielauer,
  Stöger-Pollach, and Haslinger]{U_v241_i_113594}
M.~Scheucher, T.~Schachinger, T.~Spielauer, M.~Stöger-Pollach, and
  P.~Haslinger.
\newblock Discrimination of coherent and incoherent cathodoluminescence using
  temporal photon correlations.
\newblock \emph{Ultramicroscopy}, 241:\penalty0 113594, nov 2022.
\newblock \doi{10.1016/j.ultramic.2022.113594}.

\bibitem[Kone{\v{c}}n{\'{a}} et~al.(2022)Kone{\v{c}}n{\'{a}}, Iyikanat, and
  de~Abajo]{SA_v8_i47_eabo7853}
A.~Kone{\v{c}}n{\'{a}}, F.~Iyikanat, and F.~J.~Garc{\'{\i}}a de~Abajo.
\newblock Entangling free electrons and optical excitations.
\newblock \emph{Sci. Adv.}, 8\penalty0 (47):\penalty0 eabo7853, nov 2022.
\newblock \doi{10.1126/sciadv.abo7853}.

\bibitem[Löffler et~al.(2019)Löffler, Sack, and Schachinger]{ACSA_v75_p902}
S.~Löffler, S.~Sack, and T.~Schachinger.
\newblock Elastic propagation of fast electron vortices through amorphous
  materials.
\newblock \emph{Acta Crystallogr. A}, 75\penalty0 (6):\penalty0 902--910, 2019.
\newblock \doi{10.1107/S2053273319012889}.

\end{thebibliography}

\end{document}